\documentclass[english]{emulateapj}
\usepackage{amstext}
\usepackage{amssymb}
\usepackage{graphicx}

\makeatletter

\providecommand{\tabularnewline}{\\}

\makeatother
\begin{document}

\shorttitle{Moon-moon scattering}
\shortauthors{Perets \& Payne}
\title{Formation of irregular and runaway moons/exomoons \\ through moon-moon scattering}

\author{Hagai B. Perets\altaffilmark{1} and Matthew J. Payne\altaffilmark{2}}

\email{hperets@physics.technion.ac.il}

\altaffiltext{1}{Technion - Israel Institute of Technology, Haifa 32000, Isarel}

\altaffiltext{2}{Harvard-Smithsonian Center for Astrophysics, 60 Garden St., Camridge, MA 02138, USA}

\begin{abstract}
Gas giant planets in the Solar system host large satellite systems
with multiple regular and irregular moons. Regular moons revolve around
their host planet in circular, low inclination short period orbits,
and are thought to form in-situ through coagulation processes. In
contrast, irregular moons have highly inclined (and even retrograde),
typically eccentric and long period orbits around their host planet.
Irregular moons are therefore often thought to have formed as unbound
objects in helio-centric orbits that were later captured to their
current orbits around the planet. However, such capture scenarios
require fine tuned conditions and/or encounter difficulties in producing
irregular moon populations around all gas giants. Here we study the
possibility that regular moons form in-situ outside the currently
observed regular-moon regime, and dynamically evolve through mutual
moon-moon scattering (as well as by secular evolution due to perturbations
by the Sun). We find that such evolution can excite the satellites
into high eccentricities and inclinations. We find that moons are
either ejected from the host planet to become runaway moons, or stay
bound and become prograde orbiting irregular moons with inclined and
eccentric orbits around their host planet. Ejected moons, unbound
to the planet, can later be temporarily re-captured by the host planet
even at retrograde orbits. Such moons are eventually re-ejected from
the system or collide with the planet, at least in the absence of
dissipative processes (e.g. collisions with existing bound moons,
a debris disk or through tidal interactions with the host planet),
not currently modeled. Uncaptured runaway moons may eventually be
ejected from the Solar system, or be captured into stable helio-centric
orbits and contribute to the populations of asteroidal or trans-Neptunian
objects. Such scenarios are relevant both for the gas-giant satellites
in the Solar system and for the dynamical evolution of exomoons. 
\end{abstract}

\section{Introduction}

Satellites of the giant planets in the Solar system are typically
categorized into regular and irregular moons. Regular moons orbit
their host planet in circular, low inclination short period orbits;
irregular moons have highly inclined (and even retrograde), typically
eccentric and have longer period orbits. Similar to planet formation,
the formation of satellite systems around giant planets in the Solar
system (and in exoplanetary systems) is difficult to study theoretically.
Regular satellites are thought to have formed through collisional
growth of smaller planetesimals in the circum-planetary disk, very
similar to planet formation in the coagulation model for the terrestrial
planets. Irregular satellites are typically thought to have formed
elsewhere in the Solar system (as asteroid/Kuiper-belt object like
planetesimals), and only later be captured to become moons. The various
models suggested have major difficuties in explaining the origin of
the irregular moons (\citealp[see][ for a review]{jew+07}). More
succesful recent capture models have been suggested; these involve
planet-planet interactions, and the existence of an additional third
ice-giant planet in the Solar system that have been later ejected
from it \citep{nes+14}. 

An oversimplified but useful approach used to study satellite formation
is to divide it into two stages, based on the importance of the effects
of gas and the circum-planetary disk on the growing satellites. The
first stage lasts a few Myr, until the dissipation of the gaseous
circum-planetary disk, following the dissipation of the gaseous protoplanetary
disk which fuels it. At the end of this stage large moons should have
formed \citep{can+06} in addition to and (possibly many) smaller
moonlets. 

The second stage begins following the gas dissipation. At this stage
the evolution is driven primarily by gravitational interactions and
collisions between the moons. In comparison with the extensive study
of the first stage of satellite-system formation in a gaseous disk,
the dynamical relaxation through moon-moon scattering at the later
stage have scarcely been explored. Here we study this dynamical evolution
using analytic techniques and N-body simulations of moon-moon scattering
around giant planets, and suggest that it can play a major role in
the build-up and sculpting of the architecture of satellite systems,
both in the Solar system as well as in (not yet detected) exomoon
systems. 

This paper is structured as follows. We first provide a brief overview
of the various processes seen in our N-body moon-moon scattering simulations,
which are later discussed in more detail. We introduce the N-body
simulations used to study the satellite systems, and discuss the initial
conditions following the dissipation of the gaseous circum-planetary
disk, followed by a brief discussion of the stability of multi-moon
systems. We then provide an analytic context to dynamical relaxation
of multi-moon systems. We describe several examples of multi-moon
systems studied through N-body simulations and present their evolution
and outcomes, demonstrating the possible outcomes. Finally we discuss
the evolution of multi-moon system through scattering and their implications
for satellites systems as well as the populations of minor bodies
in the Solar system (asteroids, trans-Neptunian objects, comets etc).

\section{Overview\label{sec:Overview}}

Dynamical evolution of a satellite system goes through various stages.
We will first review these stages, and then discuss them in more detail
both analytically and through examples from N-body simulations. 

\textbf{Instability}: Following the formation of the moons in an extended
circum-planetary disk, they mutually perturb each other gravitationally.
As long as a gaseous disk or a planetesimal disk exist, such disks
might dissipate the perturbation and stabilize the system. After the
disk is gone, the mutual perturbations may destabilize the systems
on short timescales, if their masses are large and/or the separations
between them are small enough (similar to the well studied planet-planet
scattering case;e.g. \citealp{rs+96}). The perturbations can then
grow and lead to orbit crossing between the various moons. Here we
study only the evolution of satellite system at the onset of dynamical
instability after the gaseous disk has been gone. 

\textbf{Scattering, collsional growth and production of prograde irregular
satellites:} In an unstable system the moons will scatter each other
gravitationally, leading to orbital changes, exciting the eccentricity
and inclinations of the satellites. When the relative velocities between
the moons become larger than the escape velocity from their surface,
physical collisions between moons will occur, which might alleviate
further growth of the moons and/or their possible disruption. Moons
excited to high eccentricities may collide with the host planet. Alternatively,
a moon can be scattered to a large distance extending beyond the stable
region around the planet (typically $\sim1/3R_{H}$ or even $\sim2/3R_{H}$
for prograde and retrograde orbits, respectively), and then be ejected
from the system to become a runaway moon. Through the scattering process
moons can migrate in the disk and exchange positions to produce a
mixed order, i.e. erasing the signature of the original moon ordering. 

\textbf{Secular evolution and the Kozai-Lidov ``barrier''}: Moons
excited to high inclinations with respect to the orbital plane of
the host planet around the Sun, can become susceptible to large amplitude
secular orbital evolution due to the tidal perturbations by the Sun
or by other moons (so called Kozai-Lidov oscillations). Such secular
evolution can drive the satellite orbits into extremely eccentric
orbits, and eventually lead to their collision with another moon or
the planet, or to their ejection from the system. It is therefore
difficult for moon scattering processes to directly lead to excitation
of retrograde orbits from initially prograde orbits. A similar limitation
on eccentricities could also be seen in planet-planet scattering simulations
of exo-planetary systems \citep[e.g. ][]{nag+08}. Scattering to stable
high inclination orbits is therefore limited by a Kozai-Lidov ``barrier''. 

\textbf{Temporary and permanent re-capture of runaway moons and the
origin of retrograde irregular moons}: Following their ejection, runaway
moons, now on helio-centric orbits, are likely to interact again with
the host planet. Some can interact with the other Solar system planets
and/or protoplanetary disk and may obtain a stable helio-centric orbit.
Others might strongly interact with the planet and even be ejected
from the Solar system. Runaway moons can also be re-captured temporarily
into orbits around the planet even into retrograde orbits with respect
to the orbit of the planet around the Sun. Such re-captured moons
can potentially dissipate some of their kinetic energy through scattering
and/or collisions with other moons orbiting the planet and/or tidal
interaction with the planet (i.e. somewhat similar to planet capture
from wide eccentric orbits into close retrograde orbits in planet
scattering simulations; \citealp{nag+08,nag+11,bea+12}), and thereby
be re-captured into permanent orbits . We conjecture that such dissipative
processes could produce retrograde irregular moons (similar to the
models envisioning capture of objects formed independently in the
protoplanetary disk). Here we only model purely dynamical interactions
and do not include dissipative processes (beside direct collisions),
which will be explored elsewhere.

\section{N-body simulations\label{sec:N-body-simulations}}

Throughout this paper we discuss the results of several N-body simulations
(using the mercury code \citealp{cha99}) of satellite systems. In
following papers we will describe the statistical results of a large
sample of simulations through the study of a wide range of initial
conditions. The current paper focuses on presenting the basic processes
and outcomes of moon-moon scattering process, and only presents results
from a small number of simulations to provide explicit examples. The
results of these specific simulations are corroborated by the much
larger set of simulations (see Payne \& Perets; paper I), for detailed
discussion of the extended N-body simulations). Table 1 provides the
the initial conditions for these simulations . We have used a single
Neptune like planet (Neptune mass and separation from the Sun), orbited
by a large number of moons (10-50 massive moons, and in some cases
up to 1000 massless particles corresponding to low mass moons). In
all simulation the outermost moon is placed at a distance ${\rm R}_{{\rm out}}=\sim1/6R_{{\rm H}}$
(i.e. at stable orbits comparable to the outermost prograde satellites
in the Solar system; ${\rm R}_{{\rm H}}$ is the planet Hill radius)
and subsequent moons are placed internal to this moon at separations
of $5$ mutual Hill radii from each other. In all simulations shown
here the moons are put on co-planar orbits coinciding with the orbital
plane of the planet orbit around the Sun; with only small random inclinations,
distributed randomly between $0^{\circ}-1^{\circ}$ and 0-0.01 eccentricities.
Though we show results of moon-moon scattering around a Neptune like
planet, results of similar satellite systems around other Solar system
like planets show similar results (Perets et al., in prep.). 

\begin{table*}
\begin{tabular}{|c|c|c|c|c|c|c|c|}
\hline 
\# & Planet & $N_{{\rm moon}}$ & MF & ${\rm N_{{\rm test}}}$ & ${\rm R_{{\rm out}}}$ & $T_{{\rm sim}}$ (yrs) & \tabularnewline
\hline 
1 & Neptune & 10 & equal mass; $m=2\times10^{25}$g & 0 & 0.17$R_{H}$ & $10^{6}$ & \tabularnewline
\hline 
2 & Neptune & 50 & equal mass; $m=5.5\times10^{25}$g & 0 & 0.17$R_{H}$ & $6.3\times10^{7}$ & \tabularnewline
\hline 
3 & Neptune & 20 & equal mass; $m=9.3\times10^{22}$ g & 1000 & 0.17$R_{H}$ & $10^{5}$ & \tabularnewline
\hline 
4 & Neptune & 3 & equal mass; $m=2\times10^{25}$ g & 0 & 0.17$R_{H}$ & $10^{7}$ & \tabularnewline
\hline 
\end{tabular}

\caption{\label{tab:Simulation-models}Simulation models}
\end{table*}

\section{Circum-planetary disk\label{sec:Circum-planetary-disk}}

The first stage of moon formation has been explored by several groups
who provided a plausible explanation for the origin of the regular
moons in the a small (typically $<$few$\times$0.01$R_{H}$) circum-planetary
disk. Studies of the formation of circum-planetary disks, however,
suggest that such disks should extend up to a large fraction of the
planetary Hill radius (up to approximately $0.33R_{H}$; \citealp[e.g. ][]{ayl+09,tan+12}),
and are not truncated at small separations, as was hitherto assumed
\citep[e.g. ][]{can+06,ogi+12}; based on older non-resolved simulations
of the gaseous disks; e.g. \citealp{lub+99}). Formation of regular
moons can therefore potentially extend much beyond the region of the
currently observed regular moons. Given these results, the likely
initial conditions for a satellite system is the formation of a multi-moon
system extending up to a large fraction of $R_{H}$. Consequently,
the initial condition assumed throughout this study are the initial
existence of a disk like population of moons extending up to 0.3 $R_{H}$
(the typical stability region for gas planet moons). A detailed study
of the \textit{initial stages }of regular moons formation in an extended
circum-planetary disk is beyond the scope of this paper and will be
discussed elsewhere. Here we focus on the \textit{dynamical evolution}
of moons in extended disks.

\section{Stability\label{sec:Stability}}

A general criteria for the stability of a multi-object Keplerian system
has not been found yet. For a system with two low mass planets on
circular, nearly co-planar orbits, a stability criteria was found
by \citet{mar+82} and \citet{gla+93}, who showed that such systems
do not have a close encounter (they are \textquoteleft{}\textquoteleft{}Hill
stable\textquoteright{}\textquoteright{}) if their semi-major axes
are separated by

\begin{eqnarray}
\Delta a & = & a_{2}-a_{1}=\nonumber \\
 & = & 2\sqrt{3}R_{H,mutual}=2\sqrt{3}\left[\frac{a_{1}+a_{2}}{2}\left(\frac{m_{1}+m_{2}}{3M_{\odot}}\right)^{1/3}\right]\text{.}\label{eq:stability}
\end{eqnarray}

We can now scale this criteria for a satellite system around a planet,
and consider two moons (where the Solar mass is replaced by the planetary
mass, and $m_{1}$, $m_{2}$ now refer to the satellite masses). For
a system of two moons with $a_{2}>a_{1}$to be stable we then require
from Eq. \ref{eq:stability} that

\begin{equation}
a_{2}>\left[\frac{1+\sqrt{3}\left(\frac{m_{1}+m_{2}}{3M_{p}}\right)^{1/3}}{1-\sqrt{3}\left(\frac{m_{1}+m_{2}}{3M_{p}}\right)^{1/3}}\right]a_{1}.\label{eq:stability-2}
\end{equation}
From this relation one can find that some of the known pairs of neighboring
regular moons in the Solar system could have been dynamically unstable,
if they were located further away than their current orbit (e.g. in
the irregular moons region), or if they were more massive then their
current measured masses. 

In higher multiplicity systems direct N-body simulations show that
initially stable systems require a few mutual Hill radii separation
between the planets in order not to destabilize on a short timescale
\citep{cha+08}; however no general stability criteria is known for
$>2$ objects orbiting a central massive object. In all the simulations
of moon scattering we initialized all moons to reside at 4-6 mutual
Hill radii from their closest neighbor.

\section{Scattering and prograde irregular moons\label{sec:Scattering-and-prograde}}

\subsection{Single strong impulsive encounters\label{sub:Single-strong-impulsive}}

Let us first consider a single encounter between two moons, occurring
at some distance $R$ from the planet, assuming a fast encounter at
the impulse approximation limit. The kick velocity to a given moon
(assumed to be smaller, for simplicity) through scattering by a moon
of mass $M_{s}$ (and radius $r_{s}$) is of the order of

\[
\Delta v=\frac{GM_{s}}{bv_{rel}},
\]
where $b$ is the impact parameter, and $v_{rel}$ is the relative
velocity between the moons, typically comparable to the velocity dispersion
in the circum-planetary satellite disk. For a heated disk, the velocity
dispersion is of the order of $\mbox{\ensuremath{\beta}}v_{kep}(R),$
where $\beta=H/R$ is the height ratio of the disk, and $v_{kep}(R)$
is the Keplerian velocity at distance $R$ from the planet (cite),
i.e. $v_{rel}=\beta\sqrt{GM_{p}/R}$. Comparing the kick velocity
gained through a close approach (comparable to a few times the satellite
radius, $r_{s},$ i.e. $b\gtrsim3r_{s}$, for which we can assume
tidal effect are negligible) to the Keplerian velocity (for a circular
orbit) around the planet at that position, 
\[
v_{Kep}=\sqrt{\frac{GM_{p}}{R}}=\frac{GM_{s}}{3r_{s}\beta\sqrt{GM_{p}/R}},
\]
we get the radius at which a mutual kick could significantly change
the satellite orbit (where we assume a constant disk height ratio,
$\beta=0.05$)

\begin{equation}
R_{sig}=3\beta\frac{M_{p}}{M_{s}}r_{s}.\label{eq:Kick}
\end{equation}

Replacing the Keplerian velocity with the escape velocity, we can
similarly get the critical radius from which moons can be ejected
from the system (though in practice, due to the tidal perturbation
by the Sun moons ejected beyond $\sim1/2R_{H}$ would already become
unbound from the system).

For scattering by the most massive observed moons for each of the
giant planets, we find that $R_{sig}=3\beta\times0.65,\,0.16,\,0.27\,0.06$
$R_{H}$, taking the satellites Ganymede, Titan, Titania and Triton
(for Jupiter, Saturn, Uranus and Neptune, respectively). It is therefore
clear that even single strong encounters with such moons can significantly
alter the orbits of companion moons and even eject them, if found
beyond $R_{sig}$, which, for reasonable $H/R$ ratios includes most
of the stable region around these planets. 

As mentioned above, no analytic criteria for instability of multi-planets
is currently known, but simulations of multi-planet systems show that
such systems evolve into strong scatterings and orbit crossing when
the mutual distance between planets are a few mutual Hill radii or
less. In our simulations we typically separate each neighboring moon
by 4 mutual Hill radii, but we find that even larger separations (5-6
mutual Hill radii) eventually lead to strong scatterings (see also
Paper I).

\subsection{Relaxation through viscous stirring, multiple encounters, and the
origin of prograde irregular moons \label{sub:Relaxation-through-viscous}}

\begin{figure}
\includegraphics[scale=0.4]{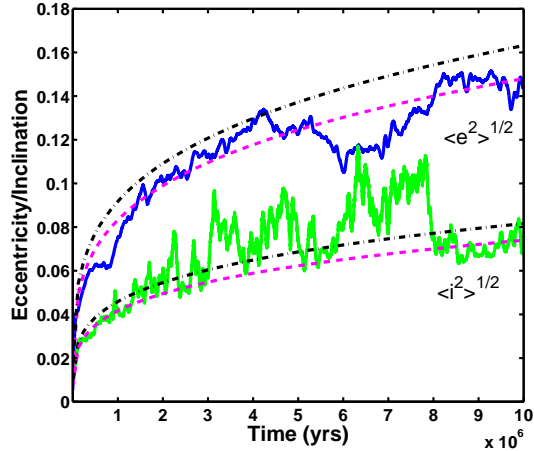}

\caption{\label{fig:viscous-stirring}Dynamical evolution of a disk of moons
around a Neptune like planet. We simulated a circum-planetary disk
composed of 50 equal mass moons ($m=3\times10^{22}$g) initially distributed
between 0.01 AU and 0.2 AU, with 4 mutual Hill radii separation between
each two neighboring moons (model 2). Shown is the evolution of the
rms eccentricity (upper solid line) and inclination (lower solid line)
compared with the analytic predictions (dash-dotted and dashed lines
correspond to $C=2$, $C=3$ respectively; see text, Eq. \ref{eq:sig-evol}).
Simulation results are smoothed with a window of 20K yrs for clarity.
The multi-moon system shows similar evolution to that of a protoplanetary
planetesimal disk. Note that at later evolutionary times (not shown
here), the disk evolution differs since the tidal perturbation from
sun become more significant for moons close to $R_{H}$ and/or at
high inclinations, where significant secular evolution may occur. }
\end{figure}

The dynamical evolution of a disk of planetesimals has been extensively
studied both numerically and analytically. Making use of similar tools
we would expect viscous stirring to excite the eccentricities and
inclinations of the moons in circum-planetary disk. For a disk composed
of $N$ equal mass moons, with mass $m$, the evolution of the velocity
dispersion in the disk, $\sigma$, with time would then go like (following
\citealp{ale+07}; see also \citealp{arm10})

\begin{equation}
\frac{d\sigma}{dt}=\frac{G^{2}Nm^{2}\ln\Lambda}{CR_{0}\Delta Rt_{orb}\sigma^{3}},\label{eq:sig-evol}
\end{equation}
for a disk (or ring) centered at radius $R_{0}$, with a radial width
$\Delta R$; $\ln\Lambda$ is the Coulomb logarithm ($\simeq9$) and
$t_{orb}$ is the Keplerian orbital period at $R_{0}$; $C$ is a
constant pre-factor ( \citealp{ale+07} find $2\leq C\le3$ in their
disk simulations). Effectively, as the disk evolves moons may be ejected
or collide and therefore $N$ and/or $m$ should evolve with time.
Nevertheless, this should serve as a good approximation at the earlier
stages of the evolution, before a significant number of ejections/collisions
of moons occur. The change in disk thickness would also affect the
maximal effective distance between encounters, thereby affecting $\Lambda$;
but the latter effect is logarithmically weak. We therefore expect
the evolution of a circum-planetary disk to be similar to that of
a protoplanetary disk, at least at the early stages and far from the
instability region where moons can be easily ejected, i.e. $\sigma(t)\propto t^{1/4}$.
In fig. \ref{fig:viscous-stirring} we show the evolution of a single
component disk in our N-body simulation (model 2 in table \ref{tab:Simulation-models}),
and compare it with the simplified analytic results (no fit; predictions
for two extreme $C$ values, $C=2,3$ are shown); we assume a disk
size of $R_{0}=\Delta R=0.1$ AU around a Neptune like (mass and distance
from the Sun) planet. We show the evolution of the rms eccentricity
and inclination ($e_{rms}=\sqrt{2}\sigma/v_{kep}$; $e_{rms}\simeq2i_{rms}$).
At later times ($>10^{7}$yrs ) the simple assumptions used in the
model break down (due to ejections/collisions which change the number
density and the mass of the moons) and the rms eccentricity of the
moons evolve slightly slower than the analytic predictions (not shown),

\subsection{Collisions and growth\label{sub:Collisions-and-growth}}

\begin{figure}
\includegraphics[scale=0.4]{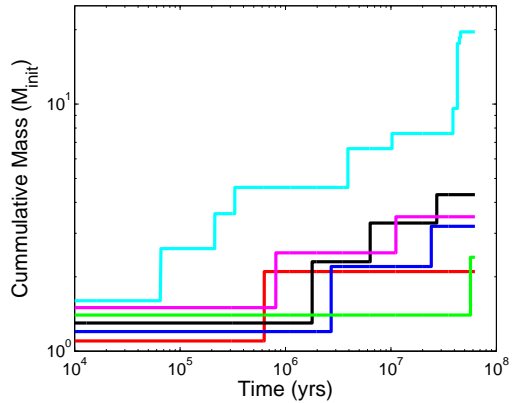}

\caption{\label{fig:mass_growth}Collisional mass growth evolution of satellites.
The mass of each satellite is shown as a function of time; all satellites
were initalized with the same mass (model 2), ${\rm M}_{init}=3\times10^{22}$
${\rm M_{\odot}}$(lines for different moons are moved up consecutively
by $0.1{\rm M_{init}}$ for better view). Of the 50 satellites at
the begining of the simulation, 24 survived to the end of the simulation
at 60 Myrs. Of these satellites the figure shows only the six which
have grown in mass through collisions. As can be seen one of these
moons has been produced through the cummulative collision of 19 satellites. }
\end{figure}

As discussed above, moon-moon scattering dynamically heats the satellite
system, leading to moon orbit crossing and potential physical collisions
and the coagulation of smaller moons into big moons. Though collisions
are included in our simulations (assuming sticking spheres model for
the collision), we focus on the late stages of satellite systems after
the main stage of the moon formation, where only a small number of
massive moons reside in the disk. A more detailed study of satellite
formation beyond the region of regular moons is beyond the scope of
this study. We note, however, that collisions do play a role in the
satellite system evolution, and a large fraction of the satellite
mutually collide. Fig. \ref{fig:mass_growth} shows an example of
the mass function evolution of a satellite system (initially with
equal mass moons). 

\subsection{Satellite system evaporation: moon ejection and runaway moons\label{sub:Satellite-system-evaporation:}}

\begin{figure}
\includegraphics[scale=0.4]{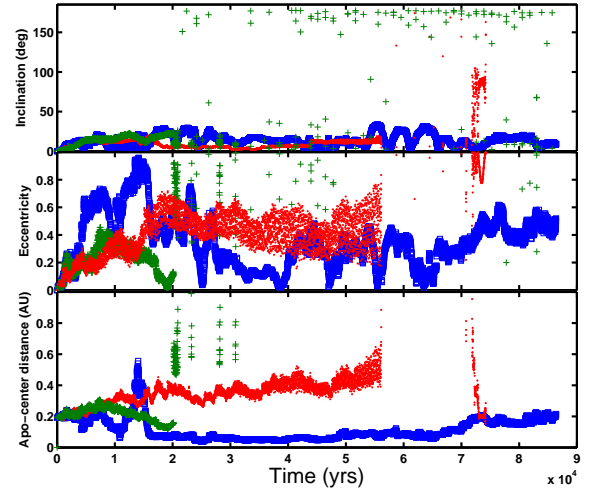}

\caption{\label{fig:orbital_evolution}Dynamical history of three different
test-mass moons in model 3, two of which are eventually ejected from
the system due to interactions with the massive satellites. Bottom:
Evolution of apo-center. The (red) dots show the slow evolution of
the satellite orbit into the unstable zone. The (green) pluses show
a satellite ejection through a strong scattering from an inner stable
orbit. Note that both ejected satellites were later temporarily captured.
The first (red dots) is recaptured and even evolves into stable small
separation orbit at very high and even retrograde inclinations (see
bottom panel), before colliding with the planet. The (blue) rectangles
show the evolution of a never ejected satellite. Middle: Evolution
of the satellite eccentricity. Note the high eccentricities at the
post-repcature stage of the satellite; eventually leading to the collsion
of the first (red dots) satellite with the planet. Top: Evolution
of satellite inclinations. Note the retrograde orbit of the first
recaptured (red dots) satellite. }
\end{figure}

\begin{figure}
\includegraphics[scale=0.4]{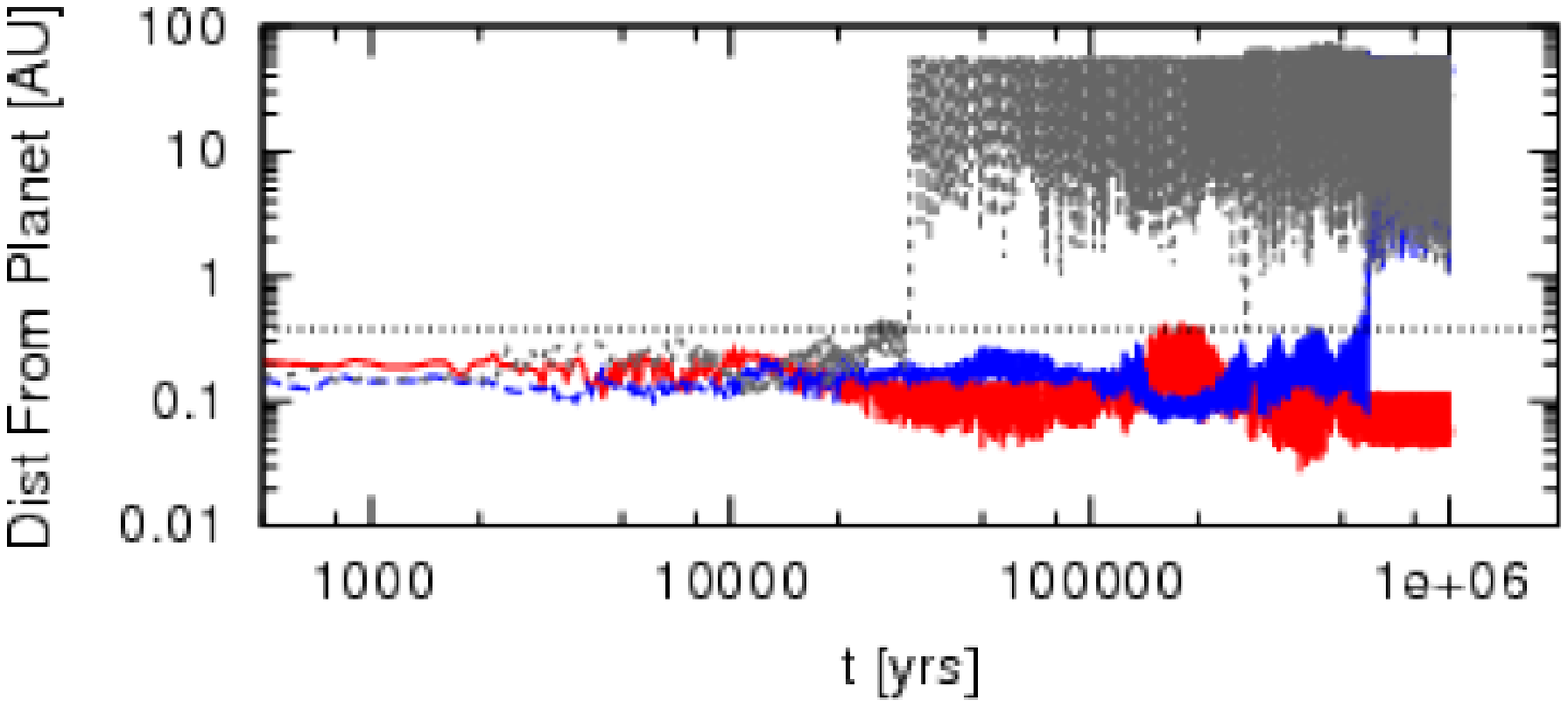}

\includegraphics[scale=0.4]{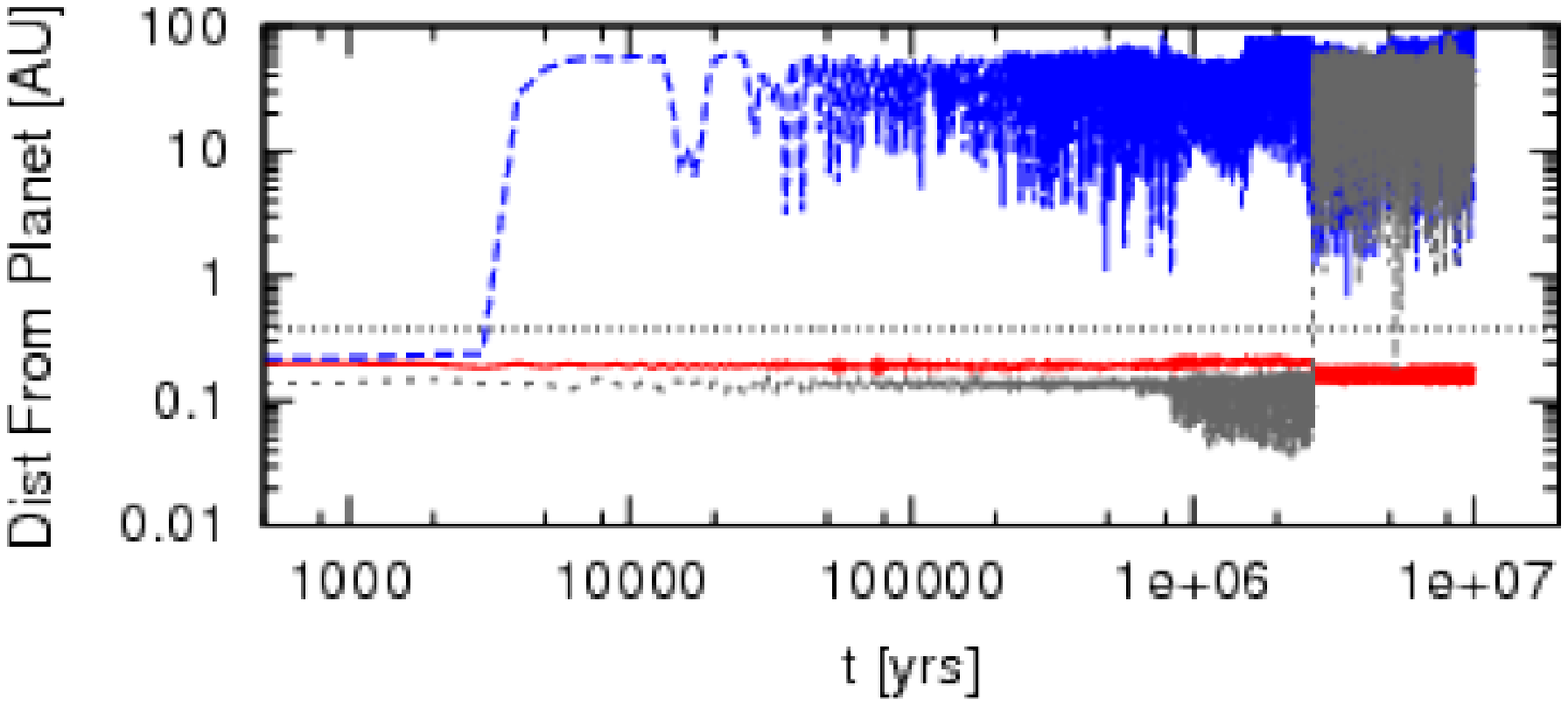}

\caption{\label{fig:orbital_evolution-2}Two examples of N-body simulation
results for the evolution of satellite system through moon-moon scattering.
The evolution captures self-excitation, scattering and ejections of
moons in a system consisting of three Triton-mass moon orbiting a
Neptune-like planet (model 4). The simulations were initialized with
three moons spaced at 3-mutual hill radii apart, with the outer-most
moon positioned at 14 Hill planetary hill radii. The separation between
the moons and the planet (in units of the planetary Hill radius) is
shown as a function of time; also plotted is a horizontal line at
0.5 Hill radii, at which to guide the reader to the approximate region
where stable satellites can not survive. The eccentricities and inclinations
of the surviving moons in such systems are excited to high eccentricities
($\sim0.2,0.5$) and inclinations ($\sim0.15$ radians). }
\end{figure}

As illustrated in Fig. \ref{eq:sig-evol} (based on results from model
2), moon-scattering excites the eccentricities (and inclinations)
of the satellites, as well as changes their semi-major axis. If the
separation of a moon from its host planet (at apocenter) extends beyond
the stability region around the planet, the tidal perturbations by
the Sun start to dominate its dynamical evolution, leading to fast
large amplitude fluctuations in its orbital elements. Typically such
evolution would, eventually, lead to the ejection of the moon from
the system as its orbit extends beyond the Hill radius. In some cases,
however, scattering by other moons close to pericenter can decrease
the apocenter distance of the moon from the planet and bring it back
into a stable orbit. We find that satellites can migrate back and
forth throughout the stability region, but, as can be seen clearly
in Figs. \ref{fig:orbital_evolution} and \ref{fig:orbital_evolution-2},
they can not survive for long beyond $\sim0.4R_{{\rm H}}$. 

The evolution of a satellite system is somewhat similar to that of
an evaporating globular cluster in the tidal field of the galaxy.
Relaxation processes lead to the contraction of the system into a
tighter configuration accompanied by the ejection of moons either
diffusively by slowly migrating beyond the stability region of the
host planet, or through more rare strong encounters, which can scatter
a moon from a stable inner orbit directly into a wide separation and
unstable orbit. Our simulations clearly demonstrate both these evaporation
channels (see Fig. \ref{fig:orbital_evolution}). 

Generally we find that, similar to multi-planet scattering results
\citep[e.g. ][]{jur+08}, at the end of our simulations typically
only 1-3 massive moons survive, irrespective of the initial number
of massive moons in the simulations; this is corroborated by a large
cohort of 3-10 massive moons simulations (using initial moon masses
comparable to those of the most massive Solar system regular moons).
The rest of the moons typically either collide with the planet or
are ejected from the system into heliocentric orbits, thereby forming
a population of ``runaway'' moons (see Fig. \ref{fig:orbital_evolution-2}
for the evolution of such runaway moons). The results of these large
phase-space studies are discussed in a companion paper (Payne et al.,
in prep.). 

\section{Secular evolution, moon recapture and retrograde irregular moons
\label{sec:Secular-evolution,-moon}}

It is well established that irregular satellites can be strongly affected
by secular Kozai-Lidov (KL) evolution due to perturbations by the
Sun \citep[see ][and references therein]{jew+07}. Such evolution
leads to periodical/quasi-periodical changes in the inclinations and
eccentricities of the satellites. The amplitude of such oscillations
can become large when the inclinations with respect to the orbit of
the host planet around the Sun are larger than $\sim40^{\circ}$(somewhat
smaller for non-zero eccentricity orbits). In particular, excitation
of large eccentricities can lead to mutual orbit crossing of the various
moons and eventually to strong scattering between them. The extremely
high eccentricities potentially reached through this process also
make the moons susceptible to physical collisions with the planet
and/or to be affected by strong tidal interactions (at pericenter).
At apocenter such moons might be ejected from the system if they approach
the instability region at apocenter (approximately 0.3-0.4 of the
planetary Hill radius).  For these reasons, moons scattered into
high (mutual) inclination (with respect to the planet orbit around
the Sun) are not likely to survive long before they collide with the
planet/other moon or are ejected from the system. Such evolution depletes
moons at high inclinations; the lifetime of moons at such inclinations
is short; they can not achieve high inclinations close to $90^{\circ}$,
which would allow them, through additional scattering (or secular
evolution) obtain retrograde orbits. Rather, they are either scattered
into at most $50^{\circ}-60^{\circ},$ and are then scattered back
to lower inclinations or they are destroyed (collision or ejection).
We note, however, that KL evolution is very sensitive to other perturbations
affecting the pericenter precession of the orbits, and can be easily
quenched due to competing interactions (e.g. the combined perturbations
of the other moons alter the pure KL evolution due to the Sun; tidal
interaction and/or the oblateness of the planet can also alter such
evolution, at least for moons achieving small pericenters; see also
\citealp{nag+11} for a similar discussion in the context of planet-planet
scattering).

Though long term stability of high inclination orbit is difficult,
we nevertheless find many moons do chive retrograde orbits throughout
their orbital evolution, sometimes even for long period of time (thousands
of years). Moon-moon scattering can therefore produce not only prograde
irregular satellites, but also retrograde irregular ones. Analysis
of their dynamical history reveals that all moons showing this behavior
have been, at earlier times, ejected from the system, and have later
been temporarily recaptured into retrograde orbits (Fig. \ref{fig:orbital_evolution}
show a typical example of such evolution). 

Such a temporary capture, if followed by a dissipative process, could
produce irregular moons; indeed this is the basic mechanism suggested
to produce irregular moons in the moon-capture scenario. Our results
therefore suggest that in-situ formation of moons followed by scattering
can lead to similar outcomes as the capture scenarios, without alluding
to capture of externally formed planetesimals, i.e. irregular moons
might be recaptured moons, in addition to, or instead of being captured
asteroids/KBOs. 

\begin{figure}
\includegraphics[scale=0.5]{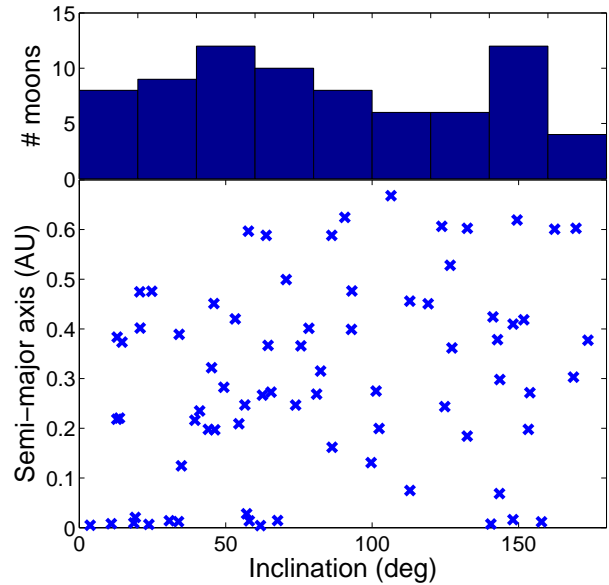}\caption{\label{fig:captured_incs}Distribution of inclinations for potentially
tidally captured/dissipated moons (in model 3; that $R_{{\rm H}}\sim0.77$
AU), at the point they where they would have (strongly) tidally interacted
with the planet (pericenter distance smaller than twice the planetary
radius). The addition of tidal interactions could potentially dissipate
sufficient energy as to recapture such moons, and quench any further
excitation of the eccentricity which would have otherwise lead to
their collision with the planet.}
\end{figure}

In our simulations we do not account for any dissipation from e.g.
interaction with gas or tidal friction during close encounters with
the host planet. Scattering by other moons can also stabilize the
orbit of such recaptured runaway moons to become permanently captured.
Nevertheless, we find no satellites that survived in retrograde orbits
more that few thousands years. In fact, we find that all the moons
which eventually collided with the host planet in our simulations
did so only after they have been ejected and had been temporarily
recaptured. In Fig. \ref{fig:captured_incs} we show the inclination
distribution of such satellites prior to their collision. This suggests
that introducing tidal interaction with the planet can have a major
part in potentially stabilizing the orbits of these moons, and potentially
produce the retrograde satellites we see today. Indeed, it had been
shown that the tidal effects are essential for producing retrograde
planets in planet-scattering simulations (e.g. Payne et al., submitted).
The potentially promising effects of adding such a dissipative process
to the moon-scattering simulations is beyond the scope of this paper
and will be discussed elsewhere.

\section{Discussion and Summary\label{sec:Discussion-and-Summary}}

The discovery of hundreds of exo-planetary systems in the last two
decades have provided new perspectives on the possible architecture
of planetary systems. In contrast to the Solar system with its co-planar
configuration of planets in nearly circular orbits, many exo-planetary
systems present planets on highly inclined (with respect to the spin
of their host star) and/or on highly eccentric orbits. Planets are
thought to form in a protoplanetary disk which could naturally account
for the existence of planets on co-planar circular orbits. The new
discoveries of eccentric/inclined planetary orbits gave rise to models
in which following the formation of multiple planets, they mutually
perturbed and scatter each other through gravitational interactions.
Such dynamical excitation eventually leads to the ejection of most
of the planets, and the survival of typically 1-3 planets on eccentric
and inclined orbits, with orbital properties consistent with observations. 

Regular moons of the Solar system gas giants have been suggested to
form in a circum-planetary disk around their host planet, in an analogue
process to the formation of the Solar system planets and exoplanets
observed in co-planar, circular orbits. As demonstrated here, massive
moons (comparable in mass to the regular moons) which formed on co-planar
circular orbits beyond the region of the regular moons can scatter
each other and/or scatter a population of low mass satellites (massless
particles in our simulations) into eccentric and inclined configurations
thereby producing irregular-like moons, serving as the satellite analogue
for eccentric/inclined exo-planetary systems. Our main results can
be summarized as follows
\begin{enumerate}
\item Most of the massive moons are ejected, typically leaving behind 1-4
surviving moons on stable configurations, irrespective of the initial
number of moons, very similar to the results of planet-planet scattering
experiments in multi-planet scattering studies (e.g. \citealp{jur+08}).
The rest of the moons are either ejected, collide with one each other
(potentially leading to further satellite growth) or collide with
the planet.
\item The surviving massive moons can obtain high eccentricities and inclinations,
similar to those of prograde irregular moons observed in the Solar
system. However, none of the massive moons obtain a stable retrograde
orbit.
\item A non-negligible fraction of the small moons (massless particles)
in our simulations ($>50$\%) are ejected from the system but are
then temporarily recaptured, for times between a few years to a few
thousand of years. Of these recaptured moons $\sim5-10$ \% have retrograde
inclinations just before they collide with the planet. 
\end{enumerate}
We therefore conclude that multiple moon systems can behave very similarly
to multiple planet systems, and can produce satellite systems with
high eccentricity and high inclination \emph{prograde} irregular satellites.
We conjecture that moons observed to temporarily attain retrograde
orbits during their evolution can potentially be stabilized through
dissipative processes (such as strong interaction with a disk/other
moons or the planet) and may also produce stable \emph{retrograde}
irregular moons, somewhat similar to the observed behavior of planet-scattering
simulations which consider tidal interactions (e.g. Payne et al.,
submitted). Note that currently no massive irregular moons beside
Triton are observed in the Solar system; though this may serve as
an important constraint on the moon-moon scattering scenario ever
happening in the Solar system, it is possible that massive irregular
moons were later removed through dissipative processes not studied
here (e.g. migration in the circum-planetary disk into close separations
to the host planet, tidal interaction with the planet, or strong disrupting
collisions). The introduction of dissipative processes in moon-moon
scattering simulations will be studied in detail elsewhere.

Finally, we note that moon-moon scattering mechanism can produce large
planetesimals in heliocentric orbits that were originally formed as
moons, and were later ejected from their host planet. Such planetesimals
might still be observed today as asteroids/Kuiper belt objects, and
may have unique kinematic and/or peculiar properties in term of composition
and/or structure compared to the background heliocentric formed planetesimal
population, due to their different origin. 

\bibliographystyle{apj}

\begin{thebibliography}{}
\expandafter\ifx\csname natexlab\endcsname\relax\def\natexlab#1{#1}\fi

\bibitem[{{Alexander} {et~al.}(2007){Alexander}, {Begelman}, \&
  {Armitage}}]{ale+07}
{Alexander}, R.~D., {Begelman}, M.~C., \& {Armitage}, P.~J. 2007, \apj, 654,
  907

\bibitem[{{Armitage}(2010)}]{arm10}
{Armitage}, P.~J. 2010, {Astrophysics of Planet Formation}

\bibitem[{{Ayliffe} \& {Bate}(2009)}]{ayl+09}
{Ayliffe}, B.~A., \& {Bate}, M.~R. 2009, \mnras, 397, 657

\bibitem[{{Beaug{\'e}} \& {Nesvorn{\'y}}(2012)}]{bea+12}
{Beaug{\'e}}, C., \& {Nesvorn{\'y}}, D. 2012, \apj, 751, 119

\bibitem[{{Canup} \& {Ward}(2006)}]{can+06}
{Canup}, R.~M., \& {Ward}, W.~R. 2006, \nat, 441, 834

\bibitem[{{Chambers}(1999)}]{cha99}
{Chambers}, J.~E. 1999, \mnras, 304, 793

\bibitem[{{Chatterjee} {et~al.}(2008){Chatterjee}, {Ford}, {Matsumura}, \&
  {Rasio}}]{cha+08}
{Chatterjee}, S., {Ford}, E.~B., {Matsumura}, S., \& {Rasio}, F.~A. 2008, \apj,
  686, 580

\bibitem[{{Gladman}(1993)}]{gla+93}
{Gladman}, B. 1993, Icarus, 106, 247

\bibitem[{{Jewitt} \& {Haghighipour}(2007)}]{jew+07}
{Jewitt}, D., \& {Haghighipour}, N. 2007, \araa, 45, 261

\bibitem[{{Juri{\'c}} \& {Tremaine}(2008)}]{jur+08}
{Juri{\'c}}, M., \& {Tremaine}, S. 2008, \apj, 686, 603

\bibitem[{{Lubow} {et~al.}(1999){Lubow}, {Seibert}, \& {Artymowicz}}]{lub+99}
{Lubow}, S.~H., {Seibert}, M., \& {Artymowicz}, P. 1999, \apj, 526, 1001

\bibitem[{{Marchal} \& {Bozis}(1982)}]{mar+82}
{Marchal}, C., \& {Bozis}, G. 1982, Celestial Mechanics, 26, 311

\bibitem[{{Nagasawa} \& {Ida}(2011)}]{nag+11}
{Nagasawa}, M., \& {Ida}, S. 2011, \apj, 742, 72

\bibitem[{{Nagasawa} {et~al.}(2008){Nagasawa}, {Ida}, \& {Bessho}}]{nag+08}
{Nagasawa}, M., {Ida}, S., \& {Bessho}, T. 2008, \apj, 678, 498

\bibitem[{{Nesvorn{\'y}} {et~al.}(2014){Nesvorn{\'y}}, {Vokrouhlick{\'y}}, \&
  {Deienno}}]{nes+14}
{Nesvorn{\'y}}, D., {Vokrouhlick{\'y}}, D., \& {Deienno}, R. 2014, \apj, 784,
  22

\bibitem[{{Ogihara} \& {Ida}(2012)}]{ogi+12}
{Ogihara}, M., \& {Ida}, S. 2012, \apj, 753, 60

\bibitem[{{Rasio} \& {Ford}(1996)}]{rs+96}
{Rasio}, F.~A., \& {Ford}, E.~B. 1996, Science, 274, 954

\bibitem[{{Tanigawa} {et~al.}(2012){Tanigawa}, {Ohtsuki}, \&
  {Machida}}]{tan+12}
{Tanigawa}, T., {Ohtsuki}, K., \& {Machida}, M.~N. 2012, \apj, 747, 47

\end{thebibliography}

\end{document}